\documentclass{aa}     
\usepackage{psfig}
\begin{document}
\thesaurus{02 (11.03.4 Coma cluster; 04.03.1; 11.03.1; 
11.04.1; 11.12.2) }
\title{Deep probing of the Coma cluster. \thanks{Based on observations 
collected at the Canada France Hawaii
telescope, operated by the National Research Council of Canada, the
Centre National de la Recherche Scientifique of France, and the University 
of Hawaii} }
\author{C.~Adami \inst{1}, R.C.~Nichol \inst{2}, 
A.~Mazure \inst{1}, F.~Durret \inst{3,4}, B.~Holden 
\inst{5}, C.~Lobo \inst{3,6} }
\institute{IGRAP, Laboratoire d'Astronomie Spatiale, 
CNRS, Traverse du
Siphon, Les Trois Lucs, F-13012 Marseille, France
\and Department of Physics, Carnegie Mellon University, 
5000 Forbes Avenue, 
Pittsburgh, PA 15213, USA
\and Institut d'Astrophysique, CNRS, Universit\'e Pierre 
et Marie Curie, 
98bis Bd Arago, F-75014 Paris, France
\and  DAEC, Observatoire de Paris, Universit\'e Paris 
VII, CNRS (UA 173),
F-92195 Meudon Cedex, France 
\and Department of Astronomy and Astrophysics, University 
of Chicago, 
5640 S. Ellis Avenue, Chicago, Illinois 60637, USA 
\and
Osservatorio Astronomico di Brera, via Brera 28, I-20121 Milano, Italy
}
\offprints{C.~Adami, adami@astrsp-mrs.fr}
\date{Received date; accepted date}

\maketitle

\markboth{Deep probing of the Coma cluster}{}

\begin{abstract}

As a continuation of our study of the faint galaxy luminosity function
in the Coma cluster of galaxies, we report here on the first
spectroscopic observations of very faint galaxies (R $\le$ 21.5) in
the direction of the core of this cluster. Out of these 34 galaxies,
only one may have a redshift consistent with Coma, all others are
background objects.  The predicted number of Coma galaxies is 6.7
$\pm$ 6.0, according to Bernstein et al. (1995, B95). If we add the 17
galaxies observed by Secker (1997), we end up with 5 galaxies
belonging to Coma, while the expected number is 16.0 $\pm$ 11.0
according to B95. Notice that these two independent surveys lead to
the same results. Although the observations and predicted values agree
within the error, such results raise into question the validity of
statistical subtraction of background objects commonly used to derive
cluster luminosity functions (e.g. B95). More spectroscopic
observations in this cluster and others are therefore urgently needed.

As a by-product, we report on the discovery of a physical structure at
a redshift z$\simeq$0.5.

\end{abstract}


\section{Introduction}

The Coma cluster of galaxies is one of the most 
intensively studied clusters 
in the sky. Its high richness and low redshift have led 
it to become 
the archetypal target for both observational and 
theoretical cluster
studies. However, one aspect of Coma that has not yet 
been fully
investigated is the distribution and dynamics of faint 
cluster
galaxies.

Indeed, we do not fully know what contribution galaxies 
fainter than
the photographic plate limit (cf. the photometry by 
Godwin et
al. 1983) make to the overall light (and mass) of Coma, 
and whether
this contribution can significantly increase the known 
baryonic mass
in the cluster. Futhermore, the exact shape of the Coma 
luminosity
function (Bernstein et al. 1995, hereafter B95, Biviano 
et al. 1995,
Lobo et al. 1997, Trentham 1997) remains in doubt, with 
suggestions
varying from a single atypical Schechter function with a 
steep
faint end slope to a Gaussian at bright magnitudes and a 
power-law at
the faint end. This confusion comes from the reliance on 
a 2D
statistical correction to subtract the background 
contribution. Most
papers analyzing cluster luminosity functions, especially 
in distant
clusters, use this method (see e.g. Driver et al. 1994, 
De Propris et
al.  1995), but such a procedure could induce severe 
correlated
errors and biases. Although the large error bars on these 
predictions do
not totally prohibit a conclusive statement on potential 
errors in the
2D statistical background subtraction, it is crucial to 
check
carefully this method.  Coma is an excellent first 
candidate for this
purpose. In the present paper, we assume H$_0$=100
km~s$^{-1}$Mpc$^{-1}$ and q$_0$=0.

\section{The sample}

\subsection{Photometry}

The photometric sample on which our study is based is one 
of the
deepest images presently available of the Coma cluster 
core taken by
B95. The data are complete to the CCD magnitude R=25 
which is far beyond the
photographic plate limit of Godwin et al. (1983).  The 
size of the
field is 7.5$\times $7.5~arcmin$^2$ and is centered on
coordinates 12h~57m~17s; 28$^{\circ }$~09' 35'' (equinox 
1950, as
hereafter). This area corresponds to the core of the Coma 
cluster near
NGC 4874 and NGC 4889.  For the spectroscopic 
observations, we have restricted 
the investigation field to the 105 faint and ``secure'' 
galaxies in the
magnitude range 18.98$\leq$R$\leq $21.5, and with a 
resolution
parameter $0.85 < \Re < 2.2$ (see B95, Fig.~3); there is 
a very low
probability to have stars or globular clusters in the 
sample according
to the B95 criteria. Note that in this range of 
magnitudes, the Coma
cluster luminosity function may rise significantly 
towards faint
magnitudes (e.g.  Lobo et al. 1997), and the 
background-to-cluster
galaxy ratio is already high ($\simeq$3.4: ratio based on 
Table 2 of B95). 
Going fainter
would only result in more background galaxies. For 99 of 
these 105
galaxies, a b$_j$ magnitude is also available.  These 
b$_j$ magnitudes
were observed at the same time as the R data but are not 
the same
signal-to-noise.

Additionally, we have checked the magnitudes of B95 by 
using the photographic
$b_{26.5}$ and $r_{24.75}$ faintest magnitudes of Godwin 
et al. (1983). 
We have found 12 common galaxies in the two samples. Even 
if the two systems
are not directly comparable, we find well constrained 
relations, which
support the B95 photometry:

b$_j$ = (0.95$\pm$0.09) $b_{26.5}$ + (0.54$\pm$0.22)

and 

R = (0.80$\pm$0.10) $r_{24.75}$ + (-1.12$\pm$0.19)

\subsection{Spectroscopy}

We have observed spectroscopically some of the targets 
using the 3.6m
CFHT with the MOS spectrograph during three nights in 
March
1997. Because of the bad weather, only half a night was 
available,
during which we have obtained spectra for 43 objects out 
of the 105
selected from the B95 total sample. Two exposures of 1 
hour have been
used for each mask. The first mask was exposed with
many intermittent and diffuse clouds and the spectra of 
one of the two
exposures were cut by a wrongly placed R filter. We 
preferentially
chose B95 objects, but we also put slits on randomly 
selected objects
with a galaxy appearance to optimize the number of slits 
per mask.

\begin{table*}
\caption[]{Table of the secure redshifts with more than 2 
lines 
(ALG or ELG) or secure with the shape of the line: 
galaxy location in arcmin with respect to the center 
12h~57m~17s; 
28$^{\circ }$~09'~35'', R and b$_j$ magnitudes, absolute 
magnitude M$_{\rm R}$, line used to determine the 
redshift 
and other available lines. The velocities obtained with a 
template analysis are also given for the two templates 
used with their Tonry \& Davis (1979) correlation 
coefficient in parentheses. The galaxies of the distant 
cluster are marked with an asterisk.}
\begin{flushleft}
\small
\begin{tabular}{ccccccccccccl}
\hline
\noalign{\smallskip}
Redshift & A/ELG & (x,y) & R & b$_j$-R & M$_{\rm R}$ & 
Line & Other 
lines & ~~~Redshift & ~~~Redshift \\
 & & (arcmin) & & & & & & ~~~~M31 & ~~~~stars \\
\hline
\noalign{\smallskip}
0.3233 $\pm$ 0.0005 & ELG & 5.02,0.87 & / & / & / & 
H$\beta$ & 
[OIII] & & \\
0.8765 $\pm$ 0.0014  & ELG & 4.69;-2.96 & / & / & / & 
[OII] &
[OIII] \\
0.5791 $\pm$ 0.0010 & ELG & 4.61;0.01 & / & / & / & [OII] 
& [OIII] & 
& \\
0.8765 $\pm$ 0.0014 & ELG & 4.15;1.35 & / & / & / & [OII] 
& [OIII] & 
& \\
0.8242 $\pm$ 0.0010 & ALG & 3.70;-0.80 & / & / & / & 
H$\&$K & & & \\
0.8294 $\pm$ 0.0007 & ALG & 3.22;-2.69 & 20.55 & 1.06 & 
-25.23 
& H$\&$K & & 0.6699 (2.2) & 0.8540 (2.4) \\
0.6773 $\pm$ 0.0010 & ELG & 2.95;1.33 & 20.78 & 1.97 & 
-21.50 
& [OII] & [OIII], 
H$\&$K? & 0.4909 (1.1) & 0.7440 (1.3) \\
0.3069 $\pm$ 0.0009 & ELG & 2.45;-2.32 & 19.85 & 1.31 & 
-20.30 & 
[OII] & [OIII], 
H$\alpha$ & & \\
0.2405 $\pm$ 0.0008 & ELG & 1.79;2.80 & 21.28 & 0.94 & 
-18.27 
& H$\alpha$ & 
[OIII] & & \\
0.7696 $\pm$ 0.0008 & ELG & 1.35;-0.55 & 20.97 & 1.25 & 
-21.68 & 
[OII] & H$\beta$ 
& & \\
0.2593 $\pm$ 0.0008 & ELG & 1.04;0.07 & 21.11 & 1.29 & 
-18.64 & 
H$\alpha$ & [OIII] 
& & \\
0.2512 $\pm$ 0.0005 & ELG & 1.37;1.92 & 20.42 & 1.36 & 
-19.24 & 
H$\alpha$ & [OII], 
H$\beta$ & & \\
0.5678 $\pm$ 0.0019 & ALG & 0.20;-0.92 & 20.45 & 2.44 & 
-21.32 & 
H$\&$K & & & \\
0.4496 $\pm$ 0.0011 & ELG & 1.26;-0.75 & 21.20 & 1.10 & 
-19.94 & 
H$\beta$ & [OIII] 
& & \\
0.6092 $\pm$ 0.0009 & ELG & 1.08;-0.22 & 21.26 & / & 
-20.71 & 
[OII] & [OIII] & & \\
0.6586 $\pm$ 0.0007 & ELG & 0.83;-1.30 & 20.98 & 0.89 & 
-21.21 & 
[OII] & [OIII] 
& & \\
0.2415 $\pm$ 0.0009 & ELG & -0.78;2.19 & 20.16 & 1.51 & 
-19.40 
& H$\alpha$ & [OIII]
 & & \\
0.4560 $\pm$ 0.0009 & ELG & 0.62;0.15 & 21.07 & 1.34 & 
-20.11 & 
[OII] & [OIII] & & \\
$*$0.5108 $\pm$ 0.0006 & ALG & -2.37;1.11 & 20.25 & 2.33 
& -21.23 
& H$\&$K & G & 0.5136 (2.4) & 0.5135 (1.7) \\
0.4440 $\pm$ 0.0008 & ELG & -0.41;2.93 & 21.02 & 2.01 & 
-20.08 
& H$\beta$ & 
[OIII], H$\alpha$? & & \\
$*$0.5209 $\pm$ 0.0012 & ALG & -2.61;2.64 & 20.43 & 2.28 
& -22.93 
& H$\&$K & 
H$\alpha$? & & \\
0.2644 $\pm$ 0.0007 & ALG & -0.54;-0.96 & 19.30 & 1.59 & 
-20.48 & 
H$\&$K & & & \\
$*$0.5000 $\pm$ 0.0100 & ALG & -2.77;1.41 & 19.77 & 1.78 
& 
-23.40 & 4000 A & 
H$\&$K & & \\
0.4266 $\pm$ 0.0003 & ELG & -0.93;2.03 & 20.97 & 1.11 & 
-20.03 
& [OII] & H$\beta$, 
[OIII] & & \\
$*$0.5188 $\pm$ 0.0020 & ALG & -3.03;0.37 & 19.48 & 1.69 
& 
-23.86 &  H$\&$K & [OII]? 
& & \\
0.1760 $\pm$ 0.0006 & ELG & -1.11;2.63 & 19.32 & 0.96 & 
-19.48 
& [OIII] & 
H$\alpha$ & & \\
0.3290 $\pm$ 0.0008 & ELG & -1.72;-1.26 & 19.65 & 1.45 & 
-20.68 & 
H$\alpha$ & 
H$\beta$ & & \\
0.6014 $\pm$ 0.0006 & ELG & -2.15;-1.56 & 20.30 & 1.03 & 
-21.63 
& H$\beta$ & 
[OIII] & & \\
$*$0.5034 $\pm$ 0.0016 & A/ELG & -2.81,0.58 & 20.45 & 
2.23 & 
-20.99 & [OII] & 
H$\alpha$, H$\&$K & 0.5194 (4.2) & 0.5187 (4.5) \\
$*$0.5104 $\pm$ 0.0025 & ALG & -3.14;1.22 & 18.99 & 2.12 
& -22.49 
& H$\&$K & & 0.5188 (5.0) & 0.5163 (3.7) \\
0.0227 $\pm$ 0.0006 & ELG & 1.66;1.02 & 21.30 & 1.72 & 
-12.89 & 
[OIII] & Shape of the line \\
0.7213 $\pm$ 0.0005 & ELG & -1.95;2.25 & 20.83 & 1.04 & 
-21.63 & [OIII] & Shape of 
the line \\
\noalign{\smallskip}
\hline
\normalsize
\end{tabular}
\end{flushleft}
\label{data3}
\end{table*}

We have corrected for the well known geometrical 
distortion of MOS
with the Geotran task in IRAF. The data reduction itself 
was made with
the ESO MIDAS classical reduction procedure package.

The redshifts were deduced by fitting gaussian profiles 
on well
identified lines. If there were emission and absorption 
lines in the
same spectrum, we selected emission lines to deduce the 
redshift. We
got 30 spectra (Table~1) with more than two emission 
lines for the
emission line galaxies (ELG hereafter), or H\&K 
absorption lines and
the 4000\AA\ break for the galaxies with only absorption 
lines (ALG
hereafter).

For four other spectra, we had only one line available. 
We could deduce a
redshift for two of them by using the shape of this line 
or the
absence of other lines (Table~1).  For the last two, we
have three possibilities. The first redshift could be
1.3284 $\pm$ 0.0017 ([OII]), 0.7854 $\pm$ 0.0013 
(H$\beta$) or
0.3224 $\pm$ 0.0010 (H$\alpha$). The second one could be
1.3108 $\pm$ 0.0022 ([OII]), 0.7719 $\pm$ 0.0017 
(H$\beta$) or
0.3124 $\pm$ 0.0013 (H$\alpha$). We do not have 
sufficient
spectral resolution to separate clearly the H and K 
absorption lines
and the [OIII]$\lambda$4959 and [OIII]$\lambda$5007 
emission lines. We
therefore use the respective mean wavelengths of these 
two sets of
lines to determine redshifts.

To test the reliability of our data, we also used the
cross-correlation task RVSAO/XCSAO in IRAF to measure radial
velocities. We selected the five best sky subtracted spectra which
exhibit absorption lines and adopted as templates two reference
spectra: M31 and a synthetic spectrum of stars, and used 50
iterations. The results are consistent with the redshifts obtained by
line identifications (see Table~1). For the best correlation
coefficient cR, the mean difference is only 2\% between the line
estimated redshift and the template estimated redshift.  The only
exception is for the galaxy at z=0.6773, with a difference of 9\%. But
in this case, cR is very low for both templates and the result is not
reliable.

The mean final success rate for extracting redshift measurements from
the spectra was 70\% for the two masks, and 76\% for the mask ``2"
observed in good conditions (for the B95 targets).  As a measure of
the quality of the B95 galaxy/star separation, we note here that
almost no globular clusters or stars were accidentally observed.

Finally, we end up with 34 redshifts (secure:30 ; secure with one
line:2 ; with several possibilities:~2; see Table~1 and Fig.~1).
Among these 34 galaxies, 29 have a magnitude (from B95) and five are
supplementary galaxies. Fig.~1 shows a typical spectrum at z=0.5034
with the [OII] (5604\AA), and H\&K (around 6000\AA) lines clearly
visible. We have 24 ELG and 10 ALG. The error on the redshift is taken
as the tenth of the FWHM of the fitted gaussian for a given line. For
the only galaxy which clearly shows emission and absorption lines, the
two estimations are consistent: 0.5034 $\pm$ 0.0016 and 0.5071 $\pm$
0.0018.

\begin{figure}
\vbox
{\psfig{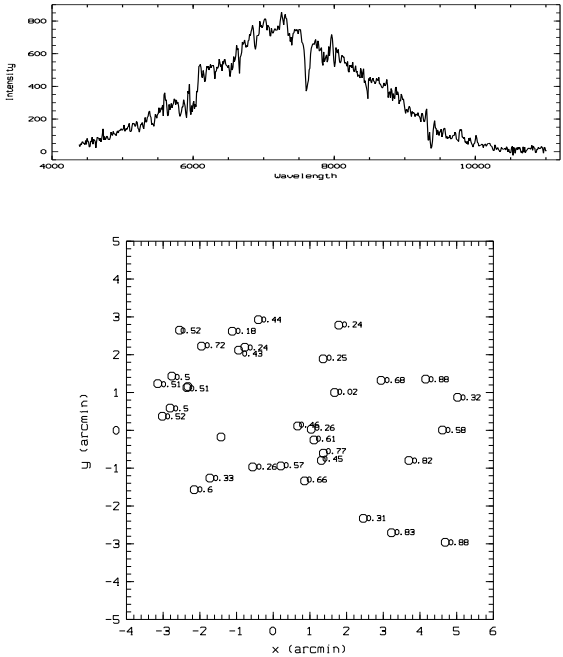}}
\caption[]{Top: typical spectrum with the [OII] and H\&K 
lines clearly 
visible (not corrected for the instrument response 
function). Bottom:
2D map of the successfully observed galaxies with the 
corresponding
redshift. The two unlabelled circles are the two galaxies 
with many 
solutions for the redshift (second unlabelled circle 
nearly superposed with
the redshift 0.51: upper left side of the figure).}
\label{}
\end{figure}

We added to our sample of 29 galaxies with a magnitude 
the 17 galaxies in 
the range of 
R magnitude [15.5;20.5] and with a (b$_j$-R) colour in 
the range [0.7;1.9]
spectroscopically observed by Secker et al. (1997, 
hereafter
S97). Interestingly, these two independent surveys lead 
to the same
results (see below).  Our total sample with redshifts is 
therefore 46 galaxies.

\section{Analysis}

\subsection{The redshift distribution observed along the 
Coma line of sight}

\begin{figure}
\vbox
{\psfig{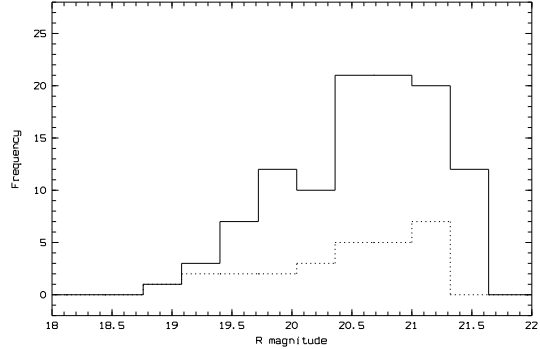}}
\caption[]{Histogram of the 105 selected galaxy 
magnitudes in the 18.98 
$\leq$ m $_R\leq$ 21.5 range (solid line) and of the 27 
successfully 
observed spectroscopically, which have only one possible 
redshift (dashed 
line) and with an available magnitude.}
\label{}
\end{figure}

The 29 successfully observed galaxies from the B95 sample 
(i.e. with an
available magnitude) are a random subsample of the global 
sample of 105 
galaxies (Fig.~2), as indicated
by the comparable magnitude distributions of the sample 
and subsample;
a Kolmogorov-Smirnov test indeed confirms that the 
probability for
these two distributions to be different is only 0.1\%. 
Hence, there is
no obvious bias in the object selection. The 32 secure 
redshifts (30 with more
than 2 lines and 2 with the shape of the line) are
almost uniformly distributed along the line of sight 
between z=0.17
and 0.88 (Fig.~3).  We potentially have only one galaxy 
in the Coma cluster at
z=0.023 and a little density peak at z$\simeq$0.5. B95 
predicted for
the [19.0;21.5] R magnitude range and for a total of 29 
galaxies (these 29 are
with a magnitude) that only
6.7 $\pm$ 6 galaxies should be in Coma (see Table~2 in 
B95). This is
consistent at 1$\sigma$ with only 1 galaxy in Coma (we do 
not take
into account the five supplementary observed galaxies 
because we do
not know their magnitudes), but the error bar is very 
large. Note that
the error on the background counts is calculated with 
robust
estimators (see B95 for details).

The B95 counts (as well as the S97 ones) predict 
9.0$\pm$5.0 galaxies
in Coma for the magnitude range covered by S97; only 4
galaxies were found to have velocities in the cluster, a 
result
comparable to that obtained with our sample.

Before merging both samples, we checked for possible 
redundant
galaxies. The fields of B95 and S97 cover the same area, 
but the
magnitude distributions are different. Only 14 out of our 
29
spectroscopically observed galaxies (with an available 
magnitude) are
in the range R [15.5;20.5]. Out of these 14, only 9 have 
a color in
the range [0.7;1.9].  Moreover, a large fraction of the 
redshifts of
S97 which are not Coma members are at z$\simeq$0.2. Among 
our 9
remaining galaxies, only 4 are at z$\leq$0.3. We 
therefore estimate
that our sample and that of S97 have a maximum of 4 
galaxies in common.

Consequently, for the 46 redshifts in this paper in the 
direction of Coma 
(obtained by putting together our data and those of S97), 
only 5 are in Coma. 
B95 would predict 16$\pm$11 in the corresponding range of
magnitudes. Once again, these numbers are consistent at 
only 1$\sigma$,
but with a very large error bar. This raises into 
question the faint end of the cluster
luminosity function derived by B95.

\begin{figure}
\vbox
{\psfig{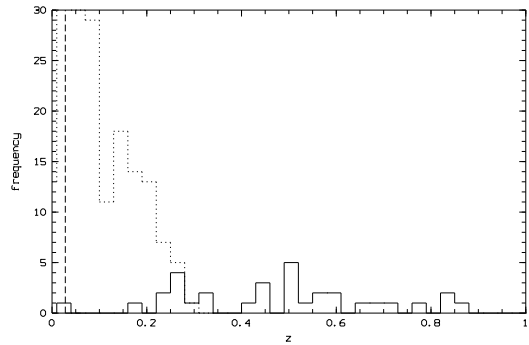}}
\caption[]{Histogram of the 32 redshifts (solid 
histogram) and of the
already available redshifts (dashed line) on the Coma 
line of
sight. The vertical long-dashed line symbolizes the 
maximum distance
for the Coma cluster assuming z$_{Coma}$=0.0232 and
$\sigma_{v,Coma}$=1000 km~s$^{-1}$.}
\label{}
\end{figure}

If we try to apply the photometric criteria based on the
color-magnitude relation, or $CMR$ (e.g. Mazure et al.  1988), to
define the cluster membership up to the magnitude limit of B95, we
obtain a puzzling result. We plot in Fig.~4 all the galaxies in the
range R [17;25] with a color available. We observe a concentration of
objects in the color range [0.85;1.30], consistent with the value
b$_j-$R=1.28 given in Frei \& Gunn (1994) for elliptical galaxies at
z=0.0. According to the color criterion, we expect that 50\% of the
galaxies with magnitudes between R=18.98 and 21.5 should be in Coma.
However, the number deduced from the spectroscopic observations is
very different, and moreover the b$_j$-R colour of the single galaxy
supposed to be in Coma is probably not in the range [0.85;1.20]
according to our colour estimates. We must therefore use this
criterion with caution. A possible explanation for this lack of faint
Coma members in this particular region of the cluster could be their
tidal disruption in the core of the cluster, as described by S97.

The existence of faint galaxies much redder than the red envelope of
the $CMR$ is important.  This could be taken as an indicator of large
metallicities for these dwarf galaxies, which are possible due to the
confinement of these systems by the intra-cluster medium (see S97).
We show in Fig.~5 the spectrum of this galaxy obtained by summing
the two exposures, together with a zoom of the only available emission
line from the first exposure with the best S/N ratio. Clearly, this
line is not H$\beta$ because we do not see H$\alpha$ at 6700\AA. The
remaining choices are [OII] and [OIII]. The shape of this line is more
consistent with the two lines of [OIII], but this is not a strong
argument. We assume then that this galaxy is in the Coma cluster and
with a b$_j-$R color equal to 1.72 (see Table~1). We have checked this
color ourselves and confirm the value of 1.72.  We note here that the
absolute magnitude of the observed Coma galaxy is $-$12.89, consistent
with faint dwarf galaxy magnitudes (e.g. Patterson \& Thuan 1996,
Stein et al.  1997). However, this redshift obviously needs to be
confirmed, and it is possible that this galaxy is also a background
object (or a very faint star), in which case there would be NO galaxy
in Coma out of the 34 objects that we observed.

We have a high percentage of ELG (70\% of the sample), but this is not
surprising, in view of the small number of galaxies in a dense
environment observed along this line of sight. Although it is possible
that we are under-estimating the fraction of galaxies in Coma at these
magnitudes, it is a well known fact that ELG occur more frequently in
the field than in clusters (see Biviano et al. 1997 for complete
references).  Moreover, the percentage of 70$\%$ is consistent with
Tresse et al.  (1996) for the CFRS survey; they found in their field
survey 85$\%$ of ELG galaxies.

\begin{figure}
\vbox
{\psfig{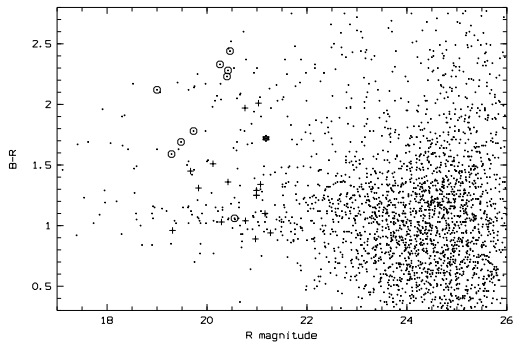}}
\caption[]{(b$_j$-R, R) distribution of the galaxies 
between R=17 and 25  
with secure redshift and color available. The galaxies 
observed 
spectroscopically are crossed, 
the ALG are circled and the only galaxy in Coma 
is the star.}
\label{}
\end{figure}

\begin{figure}
\vbox
{\psfig{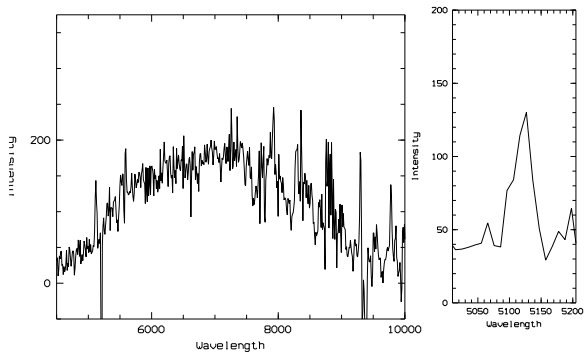}}
\caption[]{Spectrum of the galaxy supposed to be in Coma. 
Left: overall 
spectrum; right: zoom of the only available emission 
line.}
\label{toto}
\end{figure}

\subsection{A physical structure at z$\simeq$0.5 beyond 
the Coma cluster}

B95 reported the discovery of an optical subclump of 
galaxies
(approximatively at $\alpha $ = 12h~57m~14s; $\delta$ = 
28$^{\circ
}$~08'~30'') in their deep image of Coma. We have sampled 
this
subclump (see Fig.~1 and Table~1) with 6 redshifts. They 
all correspond
to the peak at z=0.5 of Fig.~3 and the spatial extension 
is 360~kpc at
z=0.5107 (mean redshift of the structure). If we assume a 
random
distribution of 34 galaxies in the (x,y,z) space where z 
is in the
[0.17;0.88] range and (x,y) is a 7.5$\times $7.5 arcmin 
square, the
probability of having 6 galaxies clustered both in 
redshift and
angularly is less than 2~10$^{-10}$.

Moreover, these 6 galaxies are ALG and represent 60\% of 
all the ALG
in the spectroscopic sample. If we keep only these 6 
galaxies with H\&K
lines (and with a template velocity when available), 
their velocity
dispersion is 660~km~s$^{-1}$ (calculated at z=0), which 
is consistent
with a cluster of galaxies, but too high for a group of 
galaxies
(e.g. Barton et al. 1996). Futhermore, this structure 
does not
correspond to any of the Rose (1977), Hickson (1982) or 
Barton et
al. (1996) groups. We note that four of these six 
galaxies have a
b$_j$-R color equal to 2.25, only moderately inconsistent 
with the
range of values calculated by Frei \& Gunn (1994) for the 
elliptical
galaxies between z=0.4 and 0.6: [2.43;2.5]. We may have 
observed the
brightest members of a distant cluster at z$\simeq $0.5 
beyond Coma.

\section{Conclusions}

As presented above, 34 faint galaxy redshifts have been 
obtained in
the direction of the Coma cluster core: 29 galaxies have 
magnitudes in
the range 19$\leq$R$\leq$21.3, and 5 do not have measured 
magnitudes.
Out of these 34 galaxies, only one may have a redshift 
consistent with the
Coma cluster, all the others being background objects. If 
we add the
S97 results, we obtain a more representative sample of 46 
redshifts
along the Coma line of sight, out of which only 5 are in 
the
cluster. Notice that these two independent surveys lead 
to the same
results.  Although the number of galaxies found to be in 
Coma is
consistent at the 1$\sigma$ level with the expected 
number of objects
in Coma predicted by B95 (16$\pm$11), such a result 
obviously raises into
question the validity of the statistical subtraction of
background objects.  Tidal disruption can be proposed as 
an
explanation of the apparent deficit of faint galaxies in 
this region of the 
core of the Coma cluster. Moreover, the presence of a 
distant cluster on the 
same line of sight probably increases the field counts. 
However, more
spectroscopic observations of very faint galaxies are 
obviously needed; we
are pursuing more data in this Coma field, but prefer to 
release the
present data to encourage other observers to also obtain 
faint
redshifts in the direction of Coma or other clusters, 
with larger telescopes 
if possible.

\begin{acknowledgements}
                          
We are indebted to Mel Ulmer and Gary Bernstein for 
making available
all the deep Coma photometry from Bernstein et al. (1995) 
and for
initially being involved in this project sometime ago.  
C.A. thanks
V. Lebrun for his help during the data reduction and for 
helpful
discussions. C.A. thanks the french GDR of Cosmology, 
CNRS, for
financial support. B.H. acknowledges the Center for 
Astrophysical
Research in Antartica, NSF OPP 89-20223, for partial 
financial
support. During this project, C.L. was fully supported by 
the
BD/2772/93RM grant attributed by JNICT, Portugal. The 
authors also
wish to thank the CFHT time allocation committee, the 
night staff and
the referee for useful comments.

\end{acknowledgements}

\vfill
\end{document}